\documentclass[aps,pra,notitlepage,twocolumn,superscriptaddress,longbibliography]{revtex4-1}

\usepackage{natbib}
\usepackage{pgfplots}
\usepackage[breaklinks]{hyperref} 
\usepackage{amssymb,bbold,bm}
\usepackage{amsmath,amsthm}
\usepackage{amsfonts,dsfont}
\usepackage{graphicx}
\usepackage{mathtools}
\usepackage{tikz}
\DeclareMathAlphabet\mathbfcal{OMS}{cmsy}{b}{n}
\usetikzlibrary{positioning}

\usepackage[normalem]{ulem}

\usepackage{color}
\usepackage{xcolor}

\newcommand{\bd}[1]{\boldsymbol{#1}}

\newcommand{\wb}{\bd{w}}

\newcommand{\Tr}{\mbox{Tr}}

\newcommand{\bra}[1]{\mbox{$\langle #1 |$}}
\newcommand{\ket}[1]{\mbox{$| #1 \rangle$}}

\definecolor{yellowm}{rgb}{0.88, 0.68, 0.13}
\definecolor{tropicalrainforest}{rgb}{0.55, 0.71, 0.0}
\definecolor{steelblue}{rgb}{0.27, 0.51, 0.71}
\definecolor{sangria}{rgb}{0.57, 0.0, 0.04}
\definecolor{patrick}{rgb}{0.14, 0.16, 0.48}

\begin{document}
\title{Excitations of Quantum Many-Body Systems via Purified Ensembles:\\ A Unitary-Coupled-Cluster-based Approach}
\author{Carlos L. Benavides-Riveros} 
\email{carlosbe@pks.mpg.de}
\affiliation{Max Planck Institute for the Physics of Complex Systems,  N\"othnitzer Str.~38, 01187, Dresden, Germany}
\affiliation{INO-CNR BEC Center, I-38123 Trento, Italy}
\author {Lipeng Chen}
\affiliation{Max Planck Institute for the Physics of Complex Systems,  N\"othnitzer Str.~38, 01187, Dresden, Germany}
\author{Christian Schilling}
\affiliation{Faculty of Physics, Arnold Sommerfeld Centre for Theoretical Physics (ASC),\\Ludwig-Maximilians-Universit{\"a}t M{\"u}nchen, Theresienstr.~37, 80333 M{\"u}nchen, Germany}
\affiliation{Munich Center for Quantum Science and Technology (MCQST), Schellingstrasse 4, 80799 M{\"u}nchen, Germany}
\author{Sebasti\'an Mantilla}
\affiliation{Max Planck Institute for the Physics of Complex Systems,  N\"othnitzer Str.~38, 01187, Dresden, Germany}
\author {Stefano Pittalis}
\affiliation{CNR-Istituto Nanoscienze, Via Campi 213A, I-41125 Modena, Italy}

\date{\today}

\begin{abstract}
State-average calculations based on mixture of states are increasingly being exploited across chemistry and physics as versatile procedures for addressing excitations of quantum many-body systems. If not too many states should need to be addressed, calculations performed on individual states are also a common option. Here we show how the two approaches can be merged into one method, dealing with a generalized yet {\em single} pure state. Implications in electronic structure calculations are discussed and for quantum computations are pointed out.
\end{abstract}

\maketitle

 Determining  properties of the excitations in quantum many-body systems is a fundamental problem across almost all  sciences. For ins\-tan\-ce, to explain the mechanism of photosynthesis \cite{Cheng2009,Cerullo2395}, human vision \cite{Johnson2015}, or photo\-vol\-taics \cite{Green2014}, one should take into account that they are main\-ly light-induced excited-state processes. The fluctuation properties of quantum spectra play also a crucial role in the characterization of quantum chaos \cite{PhysRevLett.52.1,PhysRevE.105.014109}, new states of matter \cite{PhysRevResearch.3.013143}, and more generally in the un\-ders\-tan\-ding of the temporal evolution of isolated many-body quantum systems \cite{Rigol2008}. Yet, while ground-state proper\-ties of a wide range of systems can nowadays be deter\-mi\-ned by rather accurate and computationally manageable methodologies \cite{PhysRevX.5.041041,PhysRevX.10.031058},  
 methodological developments to efficiently target excited states are highly in demand \cite{RevModPhys.74.601,SERRANOANDRES200599,doi:10.1021/acs.chemrev.8b00244,Dash2019,Gromov_2009,PhysRevLett.126.150601,Reiher2021}. 

When interested in studying the energy gaps between the ground and low-lying excited states (the optical gap and multiple neutral excitations being pro\-mi\-nent examples thereof)
we may focus on the first $M$ eigenstates $\left\{ \ket{\psi_0},\dots,\ket{\psi_{M-1}} \right\}$ of a Hamiltonian $H$
and work with
\begin{align}
\label{mixed}
\rho(\bm{w}) = \sum_{j=0}^{M-1} w_j \ket{\psi_j}\bra{\psi_j}\,.
\end{align}
The real positive weights $w_j$ are nothing else but convenient auxiliary quan\-tities.
By minimizing  the average energy $\mathcal{E}(\bm{w}) = \Tr[H \rho(\bm{w})]$ we can determine the individual states and compute all the relevant properties. The advantage of such a {\em state-ave\-ra\-ge} calculation lies in the fact that the or\-tho\-gonality of the individual states can be fulfilled automatically, but the individual states are optimal only on the average. Similarly, by mixing states with different particle numbers, calculation of electron affinities, ionization potentials, and fundamental gaps can be performed. Such ensemble calculations are increasingly being used in  traditional and emerging electronic structure methods~\cite{roos2016multiconfigurational,Methods,Kawai_2020,PhysRevResearch.1.033062,Yalouz_2021,Yalouz2022}. They are also at the center of (time-indepen\-dent) density functional \cite{PhysRevA.37.2809,PhysRevA.98.022513,Filatov2015,PhysRevLett.110.126403,Gould2020a,Gould2020,Cernatic2021,Gedeon_2021,Fromager2020} and density matrix functional~\cite{PhysRevLett.127.023001,castillo2021effective,liebert2021foundation,LS22,JL21-master} approaches to excited states.

Alternatively, {\em state-specific} calculations are also a valid option~\cite{Choo2018,Pathak2021,Lee2019,Burton2022}. But addressing states one by one, requires to satisfy  extra orthogonality conditions against previously determined states. If not particularly computationally demanding, these extra conditions can imply a {\em non-homogeneous} degradation of accuracy. When this happens, comparisons between {\em different} states get unbalanced.

Aiming at merging the advantages of state-average and state-specific calculations in one approach, here we show how a state-average calculation can be transmuted into a generalized yet {\em single} state-specific calculation.
The en\-abling idea is to map the targeted mixed state, $\rho(\bm{w})$, into a pure state
\begin{align}
\label{wfield}
\rho(\bm{w}) \rightarrow \ket{\bd{0}(\bm{w})} = \sum_j \sqrt{w_j} \ket{\psi_j}\otimes\ket{\tilde\psi_j}\,
\end{align}
belonging to a  ``double'' Hilbert space $\mathcal{H} \otimes \tilde{\mathcal{H}}$ such that 
$\mathcal{E}(\bm{w})    = \bra{ \bd{0}(\bm{w}) }  H \ket{ \bd{0}(\bm{w})}$.
When the statistical weights  are chosen as $ w_j = e^{-\beta E_j}/ \sum_j e^{-\beta E_j}$,  the state in  Eq.~(\ref{wfield}) is
the well-known {\em thermo-field}, introduced by Mat\-su\-moto and Umezawa \cite{Matsumoto,UMEZAWA}.
Central to many modern deve\-lopments in quantum sciences, this purification  of the thermal state plays an important role in quantum gra\-vi\-ty \cite{ISRAEL1976107,Maldacena_2003,maldacena2018eternal, VanRaamsdonk2010}, non-e\-qui\-li\-brium  phe\-no\-mena \cite{Lipeng2017,PhysRevLett.123.090402, Shi2020,PhysRevA.92.052116,Borrelli2021,PhysRevB.103.064309}, quantum information \cite{PhysRevB.94.155125,10.21468/SciPostPhys.6.3.034,Yu2021}, and quantum chemistry \cite{doi:10.1063/1.5089560,Harsha2019}.

As the key result presented in this Letter, we construct a {\em $\bd{w}$-field} such that 
\begin{align}
  \mathcal{E}(\bd{w})= \min_{S}  \bra{\bd{0}} e^{- S(\bd{w})} H(\bd{w}) e^{{S}(\bd{w})} \ket{\bd{0}}\,,
  \label{main1}
\end{align}
 where $\ket{\bd{0}}$ stands for  the {\em vacuum} (in the double Hilbert space),  $S(\bd{w})$ is an anti-Hermitian ``matrix'' and $H(\bd{w})$ is a unitary transformation of the Hamiltonian $H$.  
 
Crucially, we show that $S(\bd{w})$ can be fully specified via the unitary (U) coupled-cluster (CC) ansatz \cite{Taube}. This CC approach is particularly appealing because it handles the treatment of both finite (like molecules) and extended (like solids) systems \cite{PhysRevX.8.021043,10.3389/fmats.2019.00123,doi:10.1021/acs.jctc.7b00049,PhysRevB.101.241113}. Its unitary flavor can also help to solve the challenge of treating equally well dynamical and static correlations within a single approach \cite{C7CP01137G}. Yet the corresponding canonical transformation of the Hamiltonian does not truncate which makes its variational implementation not efficient on conventional classical computers. Remarkably, it was recently shown that the (Trotterized) UCC operator can be prepared at a polynomial cost on a quantum computer \cite{Peruzzo2014,RevModPhys.92.015003,PhysRevA.105.012406}.
 
Once the  minimization in Eq.~\eqref{main1} is performed, the respective eigen\-states $\ket{\psi_n}$ can be retrieved by projecting $\ket{\bd{0}(\bd{w})}$ on the non-interacting states to which $\ket{\psi_n}$ can be connected and paired to; say, $\ket{\tilde{\psi}^0_n}$.
Thereafter, taking the expectation values of the appropriate physical operators on the retrieved states, any property of individual eigenstates can be accessed. When the focus is just on energies, furthermore, we show below that extraction of the eigenstates can be avoided altogether.

In this Letter, after deriving the outlined key results,  we validate the resulting approach on a model system, and conclude touching on main perspectives.

\textit{Setting up the framework.---} 
Let us restrict ourselves to spinless fermions with non-degenerate spectrum, for simplicity. Bosonic systems may be dealt with analogously. We also assume that non-interacting states can be connected to interacting states~\footnote{In the numerical validation provided below, we explicitly handle some cases of level crossing as well.}. Inspired by the seminal idea of thermo-fields  \cite{UMEZAWA}, we invoke an auxiliary ``tilde'' space  $\tilde{\mathcal{H}}$, i.e.,~a copy of the original Hilbert space $\mathcal{H}$, such that for every state $\ket{\varphi} \in \mathcal{H}$ there is a copy $\ket{\tilde \varphi}\in \tilde{\mathcal{H}}$. For any density ma\-trix $\rho({\bd{w}}) =  \sum_j w_j \ket{\varphi_j} \bra{\varphi_j}$, with  fixed weights  $\bd{w}=(w_1,w_2,...)$, $\ket{\varphi_j}\in\mathcal{H}$ and $\bra{\varphi_i}\varphi_j\rangle = \delta_{ij}$, there is a pure state $\ket{\bd{0}(\bm{w})} = \sum_j \sqrt{w_j} \ket{\varphi_j}\otimes\ket{\tilde\varphi_j}$. 
Thus, the expectation value of any physical operator $A: \mathcal{H}\rightarrow \mathcal{H}$ can be obtained 
as:
$\bra{\bd{0}(\bd{w})} A \ket{\bd{0}(\bd{w})} = \sum_{jk}\sqrt{w_j w_k} \bra{\varphi_j} A \ket{\varphi_k} \delta_{jk} =\Tr[A \rho(\bd{w})]$,
and one  recovers the original ensemble density by tracing out all the fictitious states: $\rho(\bd{w}) = \Tr_{\tilde{\mathcal{H}}}[\ket{\bd{0}(\bd{w})}\bra{\bd{0}(\bd{w})}]$.~\footnote{It may be insightful to rewrite the $\bd{w}$-vacuum as
$\ket{\bd{0}(\bd{w})} = (\sqrt{\rho(\bd{w})}\otimes \mathds{1}) \ket{\Phi}$,
where $\ket{\Phi}\equiv \ket{\bd{0}(1,...,1)}=\sum_j \ket{\psi_j} \otimes\ket{\tilde{\psi}_j}$ is the maximally entangled state between each state and the corresponding auxiliary one \cite{Borrelli2021,PhysRevB.94.155125}.}
The field operators acting on the tilde space, i.e., the til\-de fermionic operators $\tilde{c}_m$, $\tilde{c}^\dagger_m$, obey the same anti-com\-mu\-ta\-tion rules as their untilde counterparts \cite{UMEZAWA}  and satisfy
the anti-commutation rules
$ \{c_m,\tilde{c}_m\} = \{c_m,\tilde{c}^\dagger_m\} = 0$.
By definition, operators acting on the physical space $\mathcal{H}$ do not act on states in the tilde space $\tilde{\mathcal{H}}$, and viceversa. 
	
The weighted sum of the spectrum of a Hamiltonian $H = h + W$, with $h$ and $W$ being the free (non-interacting) and the interacting Ha\-mil\-tonians respectively, can be computed by resorting to the exponential parametrization of the configuration space \cite{Helgaker2000}.  
To perform such a pa\-ra\-me\-trization let us express the corresponding eigensystems as  $H \ket{\psi_j} = E_j\ket{\psi_j}$ and $h \ket{\psi^0_j} = E^0_j\ket{\psi^0_j}$. It is well known that one may generate the states $\{\ket{\psi_j}\}$ by a uni\-ta\-ry transformation of the set $\{\ket{\psi^0_j}\}$, as they constitute another orthonormal basis of the same Hilbert space \cite{Yaris1964}. Indeed, one can write $\ket{\psi_j} = \sum_k \ket{\psi_k^0}U_{kj}$, where the coefficients are the elements of an unitary matrix $\bd{U}$. As a consequence, the eigenstates can be represented in terms of an operator transformation: 
$\ket{\psi_j} = e^{S} \ket{\psi_j^0}$,
where $S = \sum_{jk} S_{jk} \ket{\psi_j^0}\bra{\psi_k^0}$ and $S_{jk}$ is  an anti-Hermitian matrix. 
Using the freedom for the definition of the replica states in Eq.~\eqref{wfield} we fix $\ket{\tilde\psi_j} \equiv \ket{\tilde\psi_j^0}$, and obtain:
\begin{align}
\label{inter}
    \ket{\bd{0}(\bd{w})} = e^{S}    \ket{\bd{0}^0(\bd{w})}\,,
\end{align}
where $\ket{\bd{0}(\bd{w})} = \sum_j \sqrt{w_j} \ket{\psi_j} \otimes \ket{\tilde\psi^0_j}$  is the \textit{interacting} and 
 \begin{align}
 \label{free}
 \ket{\bd{0}^0(\bd{w})}= \sum_j \sqrt{w_j} \ket{\psi^0_j} \otimes \ket{\tilde\psi^0_j}    
 \end{align} 
 is the \textit{free} $\bd{w}$-field double states. 
We will write this latter state in terms of the single-particle states,  as follows:
\begin{align}
\label{wfieldfree}
 \ket{\bd{0}^0(\bd{w})} = \bigotimes^L_{m=1} \left(\sqrt{1-w_{s,m}}+ \sqrt{w_{s,m}} \,c_m^\dagger \tilde{c}_m^\dagger \right) \ket{\bd{0}}\,,
\end{align}
where $\frac{1}{2}\geq w_{s,m} > 0$ is the weight assigned to the \textit{single} mode $m$, $\ket{\bd{0}}$ is the vacuum of the double spa\-ce, and  $\bd{w}$ denotes the many-mode weights $w_{n_1,\cdots ,n_L} = \prod^L_{m=1} w_{s,m}^{n_m}(1-w_{s,m})^{1-n_m}$, with $n_m\in\{0,1\}$.
 A few  comments are in order here. First, the state $\ket{\bd{0}^0(\bd{w})}$ resembles in form the superconducting state of Bardeen-Cooper-Schrieffer theory \cite{PhysRev.108.1175}  --- but here the ``double'' occupation  is introduced for implementing the purification of mixed states rather than for describing a new phase of matter. Second, the distinctive state \eqref{wfieldfree} can efficiently be prepared on a quantum computer \cite{PhysRevResearch.4.013003}, which highlights the potential and broad scope of our proposed method.

Further information on the non-interacting case is provided in App.~\ref{appA} where,
through the operator
 $G = \sum_m \theta_m (c^\dagger_m \tilde{c}^\dagger_m - \tilde{c}_mc_m)$, with $\cos( \theta_m) = \sqrt{1-w_{s,m}}$, we  
 get
$\ket{\bd{0}^0(\bd{w})} = e^{G}\ket{\bd{0}}$ (the dependence of $G$ on the weights being understood). 
 
Going back to the interacting case, by multiplying \eqref{inter} on the left by $e^{G}e^{-G}$ we get
 \begin{align}
 \ket{\bd{0}(\bd{w})} = e^{G} e^{-G} e^{S} e^G \ket{\bd{0}}  =
 e^{G} e^{S(\bd{w})} \ket{\bd{0}}\,.
 \end{align}
Here, the short-hand notation $\mathcal{O}(\bd{w})\equiv e^{-G} \mathcal{O} e^{G}$ applies. Thus, the state-average energy of the interacting system can be expressed as a pure state expectation value
 \begin{align*}
\mathcal{E}(\bd{w})=
 \bra{ \bd{0}(\bm{w}) }  H \ket{ \bd{0}(\bm{w})} =
\bra{ \bd{0} }  e^{- S(\bd{w})} H(\bd{w})e^{ S(\bd{w})} \ket{ \bd{0}}\,.
 \end{align*}
As such, this formula resembles a single-reference calculation (in the double space) in which CC theories naturally arise \cite{Bulik2015}. This is the direction we take below.

Because the weights can be chosen in such a way that a Ritz-like variational principle for the underlying mixed states holds true \cite{PhysRevA.37.2805,parr1994density,engel2011density}, the state-average energy
 may thus be found by optimizing the anti-Hermitian matrix $S_{jk}$ that {\em minimizes}  $ \bra{ \bd{0} }  e^{- S(\bd{w})} H(\bd{w})e^{ S(\bd{w})} \ket{ \bd{0}}$, {\em as anticipated in Eq.~\eqref{main1}}. 
 Below, we give and validate an explicit form for $S$ --- please bear with us till then.
 
Thereafter,  eigenstates can be obtained
via straightforward projections:
$\ket{\psi_j} \sim \langle\tilde{\psi}^0_j\ket{\bd{0}(\bd{w})}$.
Unlike in approaches that use penalty terms to optimize the projection on {\em previously} determined states, in our approach eigenstates can be obtained {\em individually} or, in parallel, all at once. Given orthonormal single-particle orbitals to start with, formally, an optimization of a {\em single} pure (as opposed to a mi\-xed) state is all what is required!

\textit{Eigenenergies and gaps without eigenstates.---} 
Ei\-gen\-ener\-gies can be determined {\em without} having to reconstruct individual eigen\-sta\-tes. For this, let us express $\mathcal{E}(\bd{w})$ as the weighted sum of the contributions from each $N$-particle sector: $\mathcal{E}(\bd{w}) = \sum_N \mathcal{E}_N(\bd{w})$. Correspondingly, 
$\ket{\bd{0}_\varphi(\bd{w})} = \sum_{N} e^{iN\varphi} \ket{\bd{0}_N(\bd{w})}$,
where
\begin{align}
\label{eqN}
\ket{\bd{0}_N(\bd{w})} = \frac1{2\pi}\int_0^{2\pi} d\varphi \,  e^{-iN\varphi}  \ket{\bd{0}_\varphi(\bd{w})}\,.
\end{align}
Following Anderson \cite{ANDERSON19671}, we project the  BCS state into its $N$-particle components via the ``angle'' $\varphi$ (see  App.~\ref{appA}).
 
It is  convenient to ``normalize'' the $\bd{w}$-field as follows: $\ket{\bd{0}_N(\bd{w})} \rightarrow \ket{\bd{0}_N(\bd{w})} /\sqrt{\mathcal{D}(\bd{w})}$, 
with  $\mathcal{D}(\bd{w}) = (1-w_{s,1})\cdots(1-w_{s,L})$. 
As a result, the weighted sum of all the eigenenergies 
reads $\mathcal{E}_N(\bd{w}) = \sum_{\bd{n}}  w_{n_1,\cdots ,n_L} E_{n_1,...,n_L}$,
 where $w_{n_1,\cdots ,n_L}= \mu_1^{n_1}\cdots\mu_L^{n_L}$, with $\mu_{m} \equiv w_{s,m}/(1-w_{s,m})$, $n_m\in\{0,1\}$ and $\sum_m n_m = N$. 
Next, $\bd{w}'_{i}$ stands for for the tuple $\bd{w}$ where only the weight $w_{s,i} \rightarrow  w_{s,i}'$ is changed. 
 The energies of the $N$-particle 
sector can be extracted by the following  rather simple prescription:
\begin{align}
 E_{n_{i_1},...,n_{i_N}} = \frac{\Delta_{i_1} \cdots \Delta_{i_N}  \mathcal{E}_N(\bd{w})}{\prod^N_{m=1}(\mu_{i_m}-\mu_{i_m}')}\,,
 \label{eigenenergies}
\end{align}
where $\Delta_i\mathcal{E}_N(\bd{w}) \equiv \mathcal{E}_N(\bd{w}) -  \mathcal{E}_N(\bd{w}'_i)$
\footnote{ In App.~\ref{appA}, we put forward an expression using derivative w.r.t.~the weights for the non-interacting case [see Eq.~\eqref{spectrum2a}]. Along the lines of Ref.~\cite{Fromager2020}, this may be generalized to the interacting case. Numerically, however, we found more handy to use the expression above based on finite energy differences.}. In Eq.~\eqref{eigenenergies} only  variational energies $\mathcal{E}(\bd{w})$ should be considered whose  weight vectors $\bd{w}_s$, $\bd{w}_s'$ give rise to the \textit{same} ordering of ``collective'' many-body indexes $j(\bd{n})$, as explained in App.~\ref{appB}.
 
In the calculation of electron affinities $g_+$, ionization energies $g_-$, and fundamental gaps $g$, the original ground state is considered relative to the ground sates of the system with one more and one less particle. For which, we get
\begin{align}
 g_{\pm} = \sum^1_{p=0} (-1)^p\frac{\Delta_{i_1} \cdots \Delta_{i_{N\pm p}}
  \mathcal{E}_{N\pm p}(\bd{w})} {\prod^{N\pm p}_{m=1}(\mu_{i_m}-\mu_{i_m}')}\,,
\end{align}
and, thus, $g = g_{-} - g_{+}$.

\textit{Harnessing the framework.---}
Finally we exploit
the above formalism by specifying $S(\bd{w})$ 
in terms of the UCC ansatz: $S = T - T^\dagger$, where $T$ is the excitation operator defined according to $T = T_1 +  T_2 + T_3 + \cdots$ \cite{Taube}.  
Namely,
\begin{align}
     T_1 &= \sum_{mn} t_{mn} c^\dagger_m c_n, \nonumber \\
     T_2 &=  \sum_{mnrs} t_{mnrs}c^\dagger_m c^\dagger_n c_r c_s\,,
     \label{Top}
\end{align}
and higher-order terms follow the same structure where   
$m,n,r,s$  index occupied or unoccupied orbitals. Thus, $S(\bd{w}) = T(\bd{w}) -  T^\dagger(\bd{w})$. 

Due to its high accuracy for ground-state calculations, the CC ansatz is sometimes referred to as the ``gold standard of quantum chemistry'' \cite{Taube}. Although its ``unitary'' fla\-vor is 
impractical  on classical computers, it recently became clear that 
UCC wavefunction can  efficiently be handled on hybrid quantum-classical hardware like the variational eigensol\-vers \cite{Peruzzo2014,RevModPhys.92.015003}. One of the key ingredients of such an im\-ple\-men\-tation is the exact identity for each of the different UCC factors appearing in the usual Trotter formula, i.e., $\exp\left[\vartheta_{i_1\cdots i_n}^{a_1\cdots a_n}(A_{i_1\cdots i_n}^{a_1\cdots a_n} - (A_{i_1\cdots i_n}^{a_1\cdots a_n})^\dagger)\right]$, where $\vartheta_{i_1\cdots i_n}^{a_1\cdots a_n}$ are va\-ria\-tio\-nal parameters and $A_{i_1\cdots i_n}^{a_1\cdots a_n}$ are the excitation operators $ c^\dagger_{a_1}\cdots c^\dagger_{a_n} c_{i_1}\cdots c_{i_n}$ \cite{doi:10.1063/1.5133059,chen2021flexibility,D0CP01707H,PhysRevA.105.012406}. We  just need to operate the  replacements  $A_{i_1\cdots i_n}^{a_1\cdots a_n} \rightarrow A_{i_1\cdots i_n}^{a_1\cdots a_n}(\bd{w})$.

\textit{Validation.---} 
Let us consider a 1D
lattice model of spin\-less interacting fermions with Ha\-mil\-tonian:~\cite{vanNieuwenburg9269,PhysRevLett.122.040606,Morong2021}
\begin{align}
\label{hamiltonian}
    H = -\sum^L_{m=1} \left(c^\dagger_m c_{m+1} + c^\dagger_{m+1} c_{m} - U n_m n_{m+1}\right)\,.
\end{align}
Here the operator $c^\dagger_m$ ($c_m$) creates (annihilates) a fermion on lattice site $m$, and $U$ is the strength of the nearest-neighbor repulsion. For $L$ sites, the Fock space can be decomposed in $L+1$ sectors: $\mathcal{F} = \mathcal{H}_0 \oplus \cdots \oplus \mathcal{H}_L$. Only $L$ weights $w_{s,m}$ are needed in our prescription (one for each mode). 

We implemented the factorized form of UCC with singles and doubles (UCCSD). For the minimization of $\mathcal{E}(\bd{w})$ we have used the Nelder–Mead method with tolerance $10^{-5}$ \cite{Gao2012}. Increasing the number of Trotter steps beyond 4 does {\em not} substantially improve our results presented below. Further details are reported in App.~\ref{appC}.

\begin{figure}[htb]
\begin{tikzpicture}
 \node (img) {
  \includegraphics[scale=0.28]{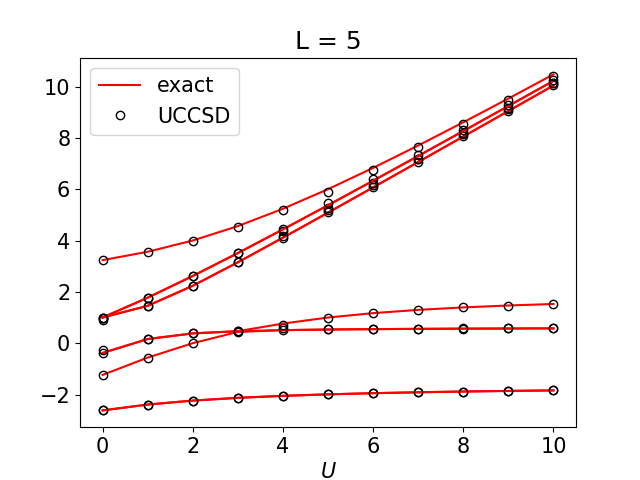}\hspace{-0.4cm}
  \includegraphics[scale=0.28]{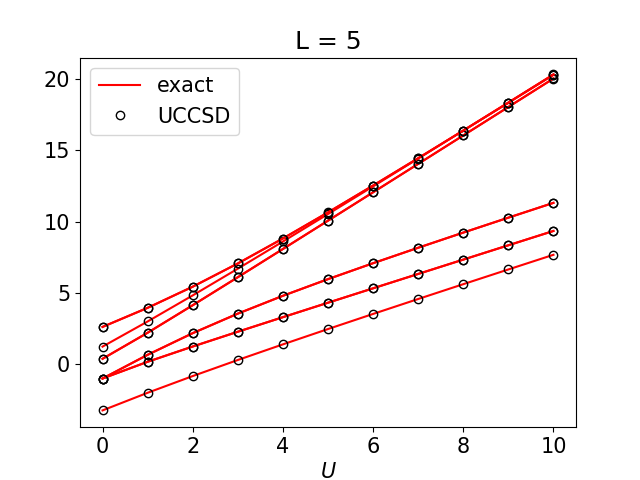}};
\node[left=of img, node distance=0cm, anchor=center, xshift=3.35cm,yshift=1.9cm,font=\color{black}] {$N = 2$};
\node[left=of img, node distance=0cm, anchor=center, xshift=7.58cm,yshift=1.9cm,font=\color{black}] {$N = 3$};
 \end{tikzpicture}
\begin{tikzpicture}
 \node (img) {\includegraphics[scale=0.28]{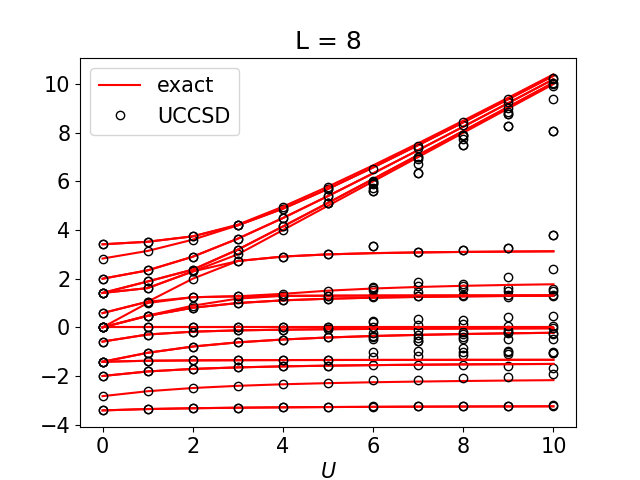}\hspace{-0.4cm}
  \includegraphics[scale=0.28]{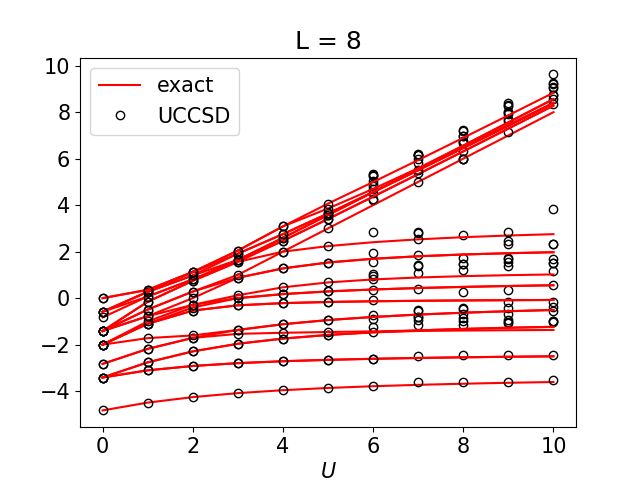}};
 \end{tikzpicture}
\caption{The spectrum of the model \eqref{hamiltonian} as a function of $U$ for $N=2,3$ and $L = 5,8$ sites. The red lines correspond to the exact energies.  Single-particle weights  $w_{s,m}$ were chosen decreasing from 0.5 with spacing $-0.5/L$. For easy of comparison, for $L = 8$ we have plotted only the lowest 25 eigenenergies.}
\label{fig2}
\end{figure}

\begin{figure}[t!]
\begin{tikzpicture}
 \node (img) {
  \includegraphics[scale=0.28]{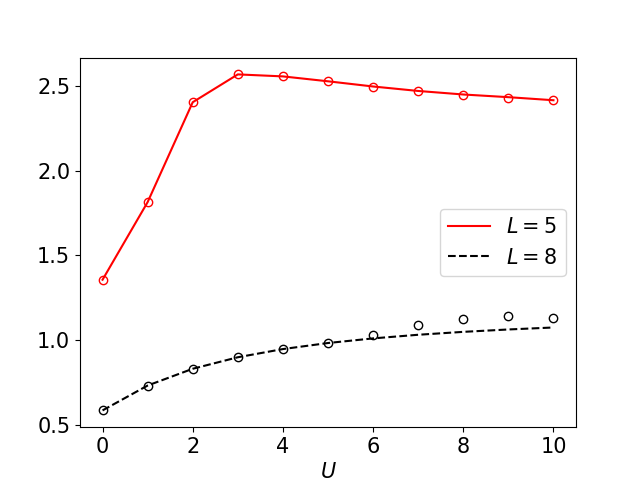}\hspace{-0.4cm}
  \includegraphics[scale=0.28]{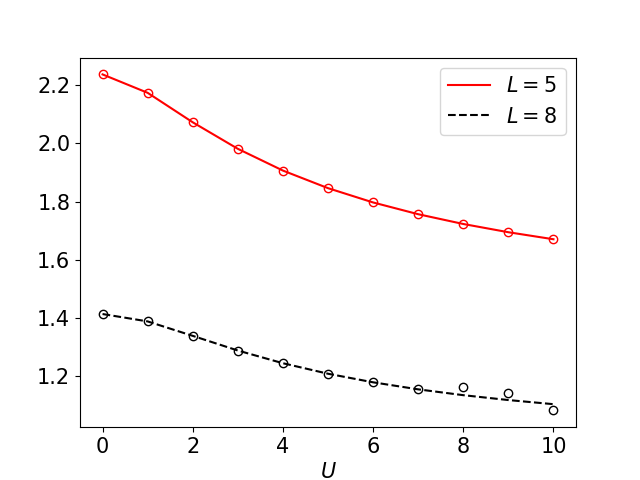}};
\node[left=of img, node distance=0cm, anchor=center, xshift=3.35cm,yshift=1.7cm,font=\color{black}] {$N = 2$};
\node[left=of img, node distance=0cm, anchor=center, xshift=7.58cm,yshift=1.7cm,font=\color{black}] {$N = 3$};
 \end{tikzpicture}
  \caption{ Neutral gaps of the model \eqref{hamiltonian} as a function of $U$ for $N=2,3$ and $L = 5,8$ sites. The solid and dashed lines correspond to the exact analytical gaps, while the red and black dots correspond to the energies computed with the energy-extraction method  \eqref{eigenenergies}. We have employed weights as in Fig.~\ref{fig2} and $w_{s,m}'= w_{s,m} + 0.005$.}
\label{gapsfig} 
\end{figure}

Let us compute the individual energies by extracting the states via the projection $\ket{\psi_j} \sim \langle\tilde{\psi}^0_j\ket{\bd{0}(\bd{w})}$ and compare them with the exact diagonalization results.  In Fig.~\ref{fig2} we  report  the energies for $L \in \{5, 8\}$. For $L=5$ our approach is near to be exact, due to reduced dimensionality of the corresponding Hilbert space. For $L = 8$, in the weakly correlation regime, the predicted energies are also in excellent agreement with the exact ones.
Discrepancies become noticeable for $U \gtrsim 6$. The source of which are mainly: (a) UCCSD cannot handle strongly interacting states, most of all; (b) the increasing number of variational parameters for large $L$ in the UCCSD ansatz (i.e., $\vartheta^{a_1a_2}_{i_1i_2}$) that must be determined in the minimization; and (c) the need of tighter tolerances in the minimization algorithm as a consequence of the  increase of the Hilbert space's dimension. Moreover, because the weights are not variational, but fixed, auxiliary parameters, the results within a given approximations may be thus conditional on those values. Analytically, we find that $\mathcal{E}(\bd{w})$ can be written as $(1-w_{s,m})\mathcal{A} + w_{s,m}\mathcal{B}$. This  linearity, however,  is observed as long as the {\em ordering} of the vector of many-body weights  does not change. Interestingly, deviation  from ({\em exact}) linearity may thus provides us with a way to sense the quality and stability of the results.

Finally, through Eq.~\eqref{eigenenergies} --- i.e., {\em without} using any of the previously extracted eigenstates ---  we determine the energy gaps between the ground state and the lowest excited state of our model system for the same sectors of Fig.~\ref{fig2}. The results shown in Fig.~\ref{gapsfig} with the exact result are, again, impressive.

\textit{Conclusions.---} 
We have  proposed a variational 
framework for determining the eigensystem of quantum many-body systems via the optimization of a {\em single} pure state. Such a pure state has the form of a {\em generalized} auxiliary thermo-field, encoding excitations rather than thermodynamics. Which, we have shown, can be determined via the unitary couple cluster (UCC) approach. Because UCC is suitable for an efficient implementation on quantum computers, our proposal may soon enable unprecedented calculations of excitations.
But the framework we have built is general and, thus, it may be exploited to gain
not only formal but also analytical and numerical advantages --- 
in any type of variational methodology for excited states based on ensemble ---
yet to be explored.

\begin{acknowledgments}
C.~L.~B.-R.~thanks Alexander Eisfeld, Frank Gro\ss mann, and Car\-lo Danieli for insightful  discussions.
 We acknowledge financial support from ``BiGmax'', the Max Planck Society’s research net\-work on big-data-driven materials science (\mbox{C.~L.~B.-R.}), the Max-Planck Gesellschaft via the MPI-PKS visitors program (L.~C.), the Deutsche Forschungsgemeinschaft (DFG, German Research Foundation), Grant SCHI 1476/1-1, the Munich Center for Quantum Science and Technology and the Munich Quantum Valley, which is supported by the Bavarian state government with funds from the Hightech Agenda Bayern Plus (C.~S.)  and from the MIUR PRIN Grant No. 2017RKWTMY (S.~P.).
\end{acknowledgments}
 
\appendix

\section{Definition and workings of the non-interacting $\bd{w}$-fields}
\label{appA}

Let us consider a {\em free} fermionic spinless Hamiltonian $h$ defined over $L$ single modes, say,
\begin{align}
    h= \sum^L_{m=1} \Omega_{m} c^\dagger_m c_m\,,
    \label{hnoni}
\end{align}
Its Fock spectrum is defined by
\begin{align}
\label{spectrum}
    h\ket{n_1,...,n_L}= E_{n_1,...,n_L} \ket{n_1,...,n_L}\,,
\end{align}
with $E_{n_1,...,n_L} = \sum_m \Omega_m n_m$ and $n_m\in \{0,1\}$, denoting the fermionic occupancies (constrained by the Pauli principle).

Following our motivation, we are then interested in the state-averaged energy
\begin{align}
\mathcal{E}(\bd{w}) &= \sum_{\bd{n}}  w_{n_1,\cdots ,n_L} E_{n_1,...,n_L} \nonumber \\
& = \sum_{\bd{n}} w_{n_1,\cdots ,n_L}  \bra{n_1,...,n_L} h \ket{n_1,...,n_L}\,,
\label{SA}
\end{align}
where $w_{n_1,\cdots ,n_L}$ must be regarded as some auxiliary parameters.
It may be useful to recall how $\mathcal{E}(\wb)$ could be determined through a variational principle. Since the modes $m$ are decoupled this is straightforward, though: The targeted ensemble quantum state $\rho$ does not contain any correlation between different modes, $\rho\equiv \rho_1 \otimes \rho_2 \otimes \ldots \otimes \rho_L$. This observation directly yields the following variant of the variational principle:
\begin{eqnarray}
\mathcal{E}(\wb) &=& \sum_{\bd{n}}  w_{n_1,\cdots ,n_L} E_{n_1,\ldots,n_L} \nonumber \\
&=& \sum_{m=1}^{L}\min_{\rho_m \in \mathcal{S}_1(w_{s,m})} \mbox{Tr}[ \Omega_{m} c^\dagger_m c_m \rho_m]\,,
\label{GOKfree}
\end{eqnarray}
where $\mathcal{S}_1(w_{s,m})$ denotes the space of density operators on the (2-dimensional) Fock space of the \emph{single} mode $m$ with spectrum $(w_{s,m},1-w_{s,m})$.

The discussion above also clarify that
\begin{align}\label{wvsws}
w_{n_1,\cdots ,n_L}(\wb_s)= \prod^L_{m=1} w_{s,m}^{n_m}(1-w_{s,m})^{1-n_m}\,,
\end{align}
where $w_{s,m}$ are
 single-mode weights (thus $s$ stands for ``single''). We chose these weights such that $\frac{1}{2}\geq w_{s,m}\geq 0$, and denote $\wb_s = \left(w_{s,0},~w_{s,1},...,~w_{s,L}\right)$. In our numerical
 experiments, in particular, we set 
 $w_{s,0} = 0.5$, $w_{s,1} = 0.5 - 0.5/L$, $w_{s,2} = 0.5 - 1/L$, $w_{s,3} = 0.5 - 1.5/L, ....>0$.

As a demonstration of the fact that $\mathcal{E}(\wb)$ is indeed a useful quantity, we note that we may directly extract the energy of specific eigenstates from
$\mathcal{E}(\wb)$  as follows:  
\begin{align}
\label{spectrum2a}
 E_{n_1,\ldots,n_L} =  \partial_{w_{s,1}^{n_1},\ldots,w_{s,L}^{n_L}}  \mathcal{E}(\bd{w}(\wb_s))\,,
\end{align}
where $\partial_{w_{s,1}^{n_1},\ldots,w_{s,L}^{n_L}}  \equiv \partial^{n_1}/\partial w_{s,1}^{n_1}+\cdots+\partial^{n_L}/\partial w_{s,L}^{n_L}$.

We next explain  how  we can express
$\mathcal{E}(\bd{w})$  in Eq.~\eqref{SA} via an expectation values taken over a {\em single}  (yet extended) pure state. For this goal, we introduce a $\bd{w}$-\textit{field} by mimicking what is done  in thermo-fields (but change the meaning and scope of the  weights) \cite{UMEZAWA}. For non-interacting modes, we set:
\begin{align}
\label{wfieldf}
 \ket{\bd{0}^0_\varphi(\bd{w})} = \bigotimes_{m=1}^L \left(\sqrt{1-w_{s,m}}+ e^{i\varphi}\sqrt{w_{s,m}} c_{m}^\dagger \tilde{c}_{m}^\dagger \right) \ket{\bd{0}}\,,
\end{align}
where $\ket{\bd{0}} = \ket{0,0,...0}\otimes\ket{0,0,...0}$ is the vacuum of the doubled Hilbert space.

Via Eq.~\eqref{wvsws},
the state in Eq.~\eqref{wfieldf} can also be written as follows
\begin{align}\label{w2}
 \ket{\bd{0}^0(\bd{w})} &=  \sum_{n_1,...,n_L} \sqrt{w_{n_1,...,n_L}} (c_{1}^\dagger \tilde{c}_{1}^\dagger)^{n_1}\cdots(c_{L}^\dagger \tilde{c}_{L}^\dagger)^{n_L} \ket{\bd{0}}\,.
\end{align}
We will come to the phase factor $e^{i\varphi}$  below in detail, let us ignored it for the time being. 
Finally, we can verify that
\begin{align}
\label{eq1}
\mathcal{E}(\bd{w}) &= \bra{\bd{0}^0(\bd{w})}h  \ket{\bd{0}^0(\bd{w})} \\
& =
\sum_{\bd{n}} w_{n_1,\cdots ,n_L}  \bra{n_1,...,n_L} h \ket{n_1,...,n_L}\,.\nonumber 
\end{align}

Next, we notice that the $\bd{w}$-field in Eq.~\eqref{wfieldf} 
can  also be written as follows:
\begin{align}
 \label{antiH}
  \ket{\bd{0}^0_\varphi(\bd{w})} = e^{G}\ket{\bd{0}}\,,
 \end{align}
where
$G = \sum_m \theta_m (e^{i\varphi} c^\dagger_m \tilde{c}^\dagger_m - e^{-i\varphi} \tilde{c}_mc_m)$ is an an\-ti-Hermitian operator and $\cos( \theta_m) = \sqrt{1-w_{s,m}}$ (dependence on the weights is understood for keeping the notation simple). In\-tro\-ducing the transformed Hamiltonian $h(\bd{w}) \equiv e^{-G} h e^{G}$, we obtain: 
\begin{align}\label{B}
\mathcal{E}(\bd{w}) &= \bra{\bd{0}}h(\bd{w}) \ket{\bd{0}}\;,
\end{align}
where the expectation values is taken over the vacuum of the double space.

Finally, let us discuss the use of the angle $\varphi \in \mathbb{R}$ in Eq.~\eqref{wfieldf}. This is the handle through which we  grab  terms with the same number $N$ of particles~\cite{ANDERSON19671}:
\begin{align}
\label{eqN}
\ket{\bd{0}^0_N(\bd{w})} = \frac1{2\pi}\int_0^{2\pi} d\varphi \,  e^{-iN\varphi}  \ket{\bd{0}^0_\varphi(\bd{w})}\,.
\end{align}
Thus, consistently with the workings of a Fourier transform,
\begin{align}
\label{Nstat}
    \ket{\bd{0}^0_\varphi(\bd{w})}  = \sum_N e^{iN\varphi} \ket{\bd{0}^0_N(\bd{w})} \,.
\end{align}

\section{Note on the Extraction of the energies of specific states}
\label{appB}

The extraction of selected excitation energies from $\mathcal{E}(\wb)$ through variation of the single-mode weights $\wb_{s}$ may requires more detailed explanation. Let us choose some single-mode weights $\wb_s$ and determine the respective global weights $\wb$ according to \eqref{wvsws}. 
As long as the  weights $w_{n_1,\ldots n_L}$ are all different,
the single-mode weight and many-mode weights are in one-to-one correspondence. Thus the various $\bd{n}\equiv(n_1,\ldots,n_L)\in\{0,1\}^L$  can be mapped  to ``collective'' labels $j \in \{1,2,\ldots,2^L\}$, which result ordered as follows
\begin{equation}
j(\bd{n})  < j(\bd{m}) \quad \Leftrightarrow \quad w_{\bd{n}} >  w_{\bd{m}}\,.
\end{equation}
In order to extract 
eigen-energies, the variational energies $\mathcal{E}(\wb(\wb_s))$ 
must be
compared to each other only for those weight vectors $\wb_s$, $\wb_s'$ that give rise to the \textit{same} ordering for $j(\bd{n})$ and $j(\bd{n'})$. In turn, this will then allow one to use the compact formula (9) in the Letter --- or Eq.~\eqref{spectrum2a} above also in the interacting case ---  to extract the energy of the  $j(\bd{n})$-th state.
In particular within an $N$ particle sector --- we have observed in our numerical experiments that --- at convergence of our UCC procedure,
 eigen-states are weighted with non-increasing weights for non-decreasing eigen-energies.

\section{Fully-polarized Fermi-Hubbard model}
\label{appC}

To test the excited-state formulation introduced in the Letter we consider a 1D lattice of interacting fully-polarized (or spinless) fermions. The system is driven by the following Hamiltonian:
\begin{align}
\label{hamil}
    H = - \sum^L_{m=1} \left(c^\dagger_m c_{m+1} + c^\dagger_{m+1} c_{m} - U n_m n_{m+1}\right)\,.
\end{align}
In the Bloch basis $c_k = \sum_m e^{imk2\pi/L}c_m/\sqrt{L}$ the Hamiltonian reads:
\begin{align*}
    H &= \sum^{L-1}_{k=0} \epsilon_k c^\dagger_k c_k \nonumber \\ &+ U_L \sum_{k_1,k_2,k_3,k_4} \delta^{Ln}_{k_1-k_2 + k_3-k_4} e^{i(k_3-k_4)2\pi/L} 
    c^\dagger_{k_1} c_{k_2} c^\dagger_{k_3} c_{k_4} \,,
\end{align*}
with $\epsilon_k = -2 \cos(2\pi k/L)$, $U_L = U/L$ and $n$ being any integer. For $L = 5$ this system can be exactly diagonalized as follows: 
\begin{itemize}
    \item The sectors $N = 0, 1$ are diagonal (in the Bloch basis) because the interaction term plays no role. 
    \item For the same reason, the sectors $N = 4$ (1 hole) and $N = 5$ (0 holes) are trivially diagonal.
    \item The sectors $N = 2$ and $N = 3$ (two holes) behave similarly (see below).
\end{itemize}
To diagonalize the sector $N = 2$ let us make use of the notation 
\begin{align}
\ket{a,b}\equiv c^\dagger_{k_a}c^\dagger_{k_b}\ket{0}\,,    
\end{align}
 and $f_{ab} \equiv e^{i(a-b)2\pi/5}$: We get (recall that $a \neq b$):
\begin{align}
\bra{a,b}H\ket{a,b} &= \epsilon_{ab} + \frac{U}5 \Bigg[4 + \sum_{c\neq \{a,b\}} \left(f_{ca} + f_{cb}\right)\Bigg]\,.
\end{align}
To evaluate the remaining non-diagonal terms, we note that for any pair $(a,b)$ there is only one pair $(c,d)$ for which $\delta^{5n}_{k_a+k_b - k_c-k_d}$ is non-zero: for $(0,1)$ the other pair is $(2,4)$; for $(0,2)$ is $(3,4)$; for $(0,3)$ is $(1,2)$; for $(0,4)$ is $(1,3)$, and for $(1,4)$ is $(2,3)$. Save a sign, the result is the same for all these pairs of pairs, namely: 
\begin{align}
\bra{a,b}H\ket{c,d} &= \delta^{5n}_{k_a+k_b - k_c-k_d}
U\,\frac{\left(f_{ac} + f_{bd}-f_{bc}-f_{ad}\right)}5\nonumber \\ &=
\pm \delta^{5n}_{k_a+k_b - k_c-k_d} U R\,,
\label{gs}
\end{align}
where $R = 2[\cos(2 \pi/5) - \cos(4 \pi /5)]/5 \approx 0.4472$.
The global sign in Eq.~\eqref{gs} depends on the order of the numbers $a,b,c,d$.

As a consequence of this  calculation, it is clear that the Hamiltonian is block diagonal with five $2 \times 2$ matrices on the diagonal. The relevant matrices are always of the form:
\begin{align}
H_{ab,cd} = \begin{pmatrix}
\epsilon_{ab} & 0 \\
0 & \epsilon_{cd}
\end{pmatrix} + U \begin{pmatrix}
 D & \pm  R \\
\pm R & 1- D
\end{pmatrix}\,,
\end{align}
where 
$D = 2[2 + \cos(2 \pi/5) + 2 \cos(4 \pi /5)]/5 \approx 0.2763$. Thus, all eigenvalues read:
\begin{align}
E^{\pm}_{ab,cd}(U) = \frac{\tau_{abcd} \pm \sqrt{\tau_{abcd}^2-4\Delta_{ab,cd}}}2\,,
\end{align}
where $\tau_{abcd}=\epsilon_a+\epsilon_b+\epsilon_c+\epsilon_d+U$ and $\Delta_{ab,cd}=(\epsilon_{ab}+ D U)(\epsilon_{cd}+ (1-D)U)-R^2U^2$.  For instance, the ground state energy is: $E_{\rm gs}(U) = (\tau - \sqrt{\tau^2-4\Delta})/2$, with: $\tau = \epsilon_{01}+\epsilon_{24}+U$ and $\Delta =(\epsilon_{01}+ D U )(\epsilon_{24}+ (1-D)U)-R^2U^2$. This calculation gives the exact eigenenergies for the sector $N =2$ that we use to benchmark our calculations in the paper. The eigenstates are straightforward to compute.

\begin{figure}[t!]
\begin{tikzpicture}
 \node (img) {\includegraphics[scale=0.24]{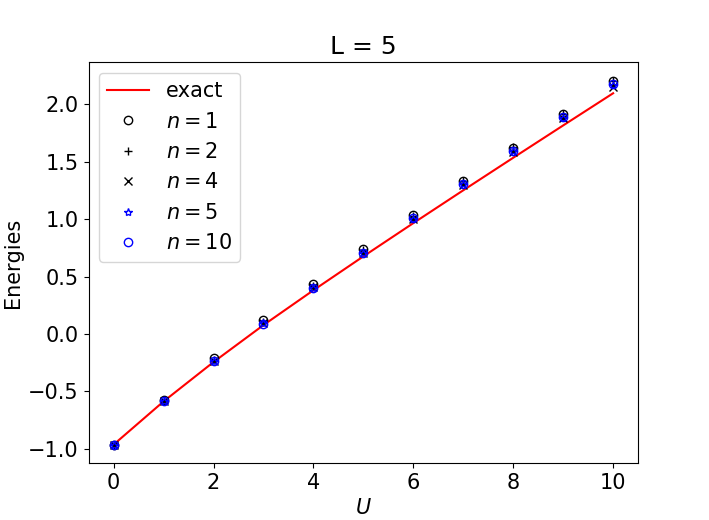}\hspace{-0.2cm} \includegraphics[scale=0.24]{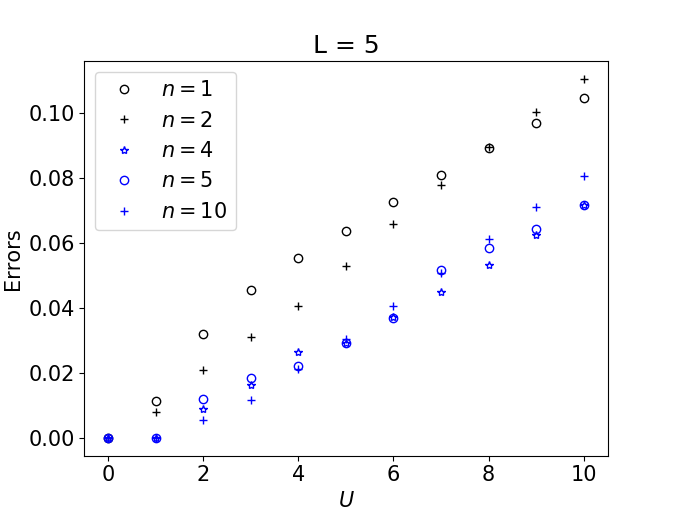}};
 \end{tikzpicture}
  \caption{The function $\mathcal{E}(\bd{w})$ for the Fermi-Hubbard model \eqref{hamil} with $L = 5$ is plotted as a function of $U$ and the number of Trotter steps (we use $\bd{w}_s = (0.5,0.4,0.3,0.2,0.1)$). Right panel: the exact values (red solid line) are compared with the variational computation using UCCSD. Left panel: deviation of the predicted weighted energies $\mathcal{E}(\bd{w})$ w.r.t.~the exact values for different Trotter step $n$.}
\label{fig2} 
\end{figure}

For $N=3$ the reasoning is essentially the same and we refrain to reproduce it here. The spectrum of these two sectors are plotted Fig.~1 of the Letter.
\\

\section{UCCSD calculations}
\label{appC} 

 The cluster operator with singles and doubles we implemented in our calculations is the following:
\begin{align*}
    U(\bd{\vartheta}) = \left[\prod_{ij\neq kl}e^{\vartheta_{ijkl}(c^\dagger_ic^\dagger_jc_kc_l-c^\dagger_lc^\dagger_kc_jc_i)} \prod_{i\neq j} e^{\vartheta_{ij}(c^\dagger_ic_j-c_j^\dagger c_i)}\right]^n\,.
\end{align*}
In this formula $n$ denotes the number of Trotter steps. We use generalized indices, i.e., $i,j,k,l$ denote occupied and unoccupied orbitals. Each term can be easily implemented as follows \cite{chen2021flexibility}:
\begin{align}
  &e^{\vartheta_{mnrs}(c^\dagger_mc^\dagger_nc_rc_s-c^\dagger_sc^\dagger_rc_nc_m)} \nonumber \\
  &\quad = 1+ 
  \sin(\vartheta_{mnrs})[c^\dagger_mc^\dagger_nc_rc_s-c^\dagger_sc^\dagger_rc_nc_m]  \nonumber \\
  &\qquad + [\cos(\vartheta_{mnrs})-1](c^\dagger_mc^\dagger_nc_rc_s-c^\dagger_sc^\dagger_rc_nc_m)^2\,,
\end{align}
and similarly for other excitations.

The function to be minimized is thus 
\begin{align}
\mathcal{E}(\bd{w},\bd{\vartheta}) =   \bra{\bd{0}}U(\bd{\vartheta},\bd{w})H(\bd{w}) U(\bd{\vartheta},\bd{w})\ket{\bd{0}}\,,
\label{loss}
\end{align}
with $U(\bd{\vartheta},\bd{w}) = e^{-G}U(\bd{\vartheta}) e^G$. To perform the minimization we employ the Nel\-der–Mead algorithm with tolerance $10^{-5}$ \cite{Gao2012}.
Notice that treating large systems has as a consequence the increase of the number of parameters that should be found in the minimization process of the loss function  \eqref{loss}. To give an idea of this point, consider the following:  while for $L =5$ there are 15 parameters for the double excitations and 10 parameters for the single excitations, for $L = 8$ these numbers jump to 210 and 28, respectively. However, it is important to note that dealing with an increasing number of parameters in a fully unitary scheme is one of the tasks that quantum computation aims at solving by mapping those unitaries to quantum logic gates.

We compute the individual energies by extracting the states via the projection $\ket{\psi_j} \sim \langle\tilde{\psi}^0_j\ket{\bd{0}(\bd{w})}$. We explored  cases from $L = 5$ up $8$ and $N = 2$ and  $N =3$ particles. In the Letter we presented the results for $L = 5,8$.

For completeness, in Fig.~\ref{fig2} we present the errors of the calculation of  $\mathcal{E}(\bd{w})$ for $L = 5$. It shows the differences between the predicted values of the function $\mathcal{E}(\bd{w})$ and the exact values. It also confirms that few Trotter steps lead to larger deviations and. Yet, we note that  $n = 4$ is an excellent approximation  in the cases we discuss.

\bibliography{Refs}

\end{document}